\documentstyle[12pt,fleqn]{article}

\def\eg{E_{\rm g}}

\def\soc{{\rm C}_{60}}
\def\rug{{\rm C}_{70}}
\def\beeq{\begin{equation}}
\def\eneq{\end{equation}}
\def\beeqa{\begin{eqnarray}}
\def\eneqa{\end{eqnarray}}

\addtocounter{section}{-1}
\begin{document}

\begin{center}

{\large {\bf{Dimerization structures\\
on the metallic and semiconducting fullerene tubules\\
} } }

\vspace{1cm}

{\rm Kikuo Harigaya$^*$}\\

\vspace{1cm}

{\sl Fundamental Physics Section, Physical Science Division,\\
Electrotechnical Laboratory,\\
Umezono 1-1-4, Tsukuba, Ibaraki 305, Japan\\
and\\
Department of Physics, University of Sheffield,\\
Sheffield S3 7RH, United Kingdom}
\vspace{1cm}

{\rm Mitsutaka Fujita}\\

\vspace{1cm}
{\sl Institute of Materials Science, University of Tsukuba,\\
Tsukuba, Ibaraki 305, Japan}

\vspace{1cm}

(Received~~~~~~~~~~~~~~~~~~~~~~~~~~~~~~~~~~~)
\end{center}

%\maketitle

\Roman{table}
%\large

\vspace{1cm}

\noindent
{\bf ABSTRACT}

\noindent
Possible dimerization patterns and electronic structures in fullerene
tubules as the one-dimensional $\pi$-con\-ju\-gated
systems are studied with the extended
Su-Schrieffer-Heeger model.  We assume various lattice geometries,
including chiral and achiral tubules.  The model is solved for
the half-filling case of $\pi$-electrons.
(1) When the undimerized systems do not have a gap, the Kekul\'{e}
structures prone to occur.  The energy gap is of the order of the
room temperatures at most and metallic properties would be expected.
(2) If the undimerized systems have a large gap ($\sim$ 1eV), the
most stable structures are the chain-like distortions where the
direction of the arranged {\sl trans}-polyacetylene chains is along
almost the tubular axis.  The electronic structures are of
semiconductors due to the large gap.

{}~

\noindent
PACS numbers: 61.65.+d, 71.20.Hk, 31.20.Pv, 71.38.+i

\pagebreak

%%%%%%%%%%%%%%%%%%%%%%%%%%  input document  %%%%%%%%%%%%%%%%%%%%%%%%%%%

\section{INTRODUCTION}

Recently, a new form of carbons, ``fullerene tubules", has been
synthesized [1,2].  A tubule has the structure like a cylinder made
from a graphite sheet.  Usually, tubules are observed in multilayered
structures.  Several number of tubules are penetrated mutually.
Distance between nearest tubular sheets is about 0.34nm and is near to
that of the separation between the layers in the graphite.  The
typical diameter of the tubules is of the order of one nanometer.
The maximum length of the tubules is more than one micrometer.
Thus, a tubule can be regarded as a new class of one-dimensional
materials.

The electronical properties have been theoretically calculated [3-7].
A tubule can show metallic or semiconductor-like properties, depending
upon its geometry and diameter [4-6].  In these studies [4-6], all of the
carbon atoms have been assumed to be equivalent.  All the bonds
with the equivalent geometry have the identical length.
The possibility of the bond alternation patterns
(namely the dimerizations) has been considered briefly concerning with
the in-plane [3,7] and out-of-plane [5] dimerizations.  They have concluded
that the strength of dimerizations might be quite small even if they
occur.

It is, however, well known that the one-dimensional systems are
unstable with respect to Peierls distortions
if a weak electron-phonon interaction is present.
Dimerizations will appear in fullerene tubules due to the Peierls
instabilities.  The dimerization patterns will depend on the structure of the
tubules.  The possible patterns have not been investigated in detail
in the previous works [3,5,7].
Therefore, it would be interesting to study what patterns
can appear when we regard tubules as quasi one-dimensional
$\pi$-conjugated systems [8].

In this paper, we extend the Su-Schrieffer-Heeger (SSH) model [9]
of conjugated polymers to the honeycomb network system
in order to apply to fullerene tubules.
The electronic properties and the possible dimerization patterns
are analyzed using the finite-size scaling method [7].
We obtain the ``Kekul\'{e} structure" for the metallic
tubules, and the ``chain-like distortion" for the semiconducting tubules
as the most stable state.  The Kekul\'{e} structure is a network
of hexagons with the alternating short and long bonds like in the
classical benzene molecule.  The chain-like distortion is a pattern
where {\sl trans}-polyacetylene chains are connected by long bonds
in the direction perpendicular to the chains.
The Kekul\'{e} or chain-like patterns will be important as fluctuations,
even though the amplitude of dimerizations is of the same order
as that of the fluctuations and therefore it is difficult to observe
the static distortions.

In the next section, we explain our model and ideas of the investigation.
In Sec. III, the Kekul\'{e} structures in the metallic tubules are
reported.  In Sec. IV, the chain-like distortions in the semiconducting
tubules are discussed.  We close this paper with several remarks
in Sec. V.

\section{MODEL AND FORMALISM}

We use the extended SSH Hamiltonian [7,9] for investigation of dimerization
patterns.  The model is
\beeq
H = \sum_{\langle i,j \rangle, \sigma} ( - t_0 + \alpha y_{i,j} )
( c_{i,\sigma}^\dagger c_{j,\sigma} + {\rm H.c.} )
+ \frac{K}{2} \sum_{\langle i,j \rangle} y_{i,j}^2,
\eneq
where $c_{i,\sigma}$ is an annihilation operator of a $\pi$-electron, the
quantity $t_0$ is the hopping integral of the ideal undimerized system,
$\alpha$ is the electron-phonon coupling, $y_{i,j}$ indicates the bond
variable which measures the length change of the bond between the $i$- and
$j$-th sites from that of the undimerized system, the sum is
taken over nearest neighbor pairs $\langle i j \rangle$.   The
second term is the elastic energy of the lattice; and the
quantity $K$ is the spring constant.  This model is numerically solved by
the iteration method used in the previous studies [7,10] under the
assumption of the adiabatic approximation.

We shall look at the dependences on parameters.  The value of $t_0 = 2.5$eV
is taken from that of graphite and polyacetylene, and has been used
in the previous papers [7,10].
We fix the spring constant $K=49.7$eV/\AA$^2$
and change the coupling constant $\alpha$, so that the dimensionless
electron-phonon coupling $\lambda \equiv 2\pi \alpha^2 / \pi K t_0$,
analogous to that in BCS superconductivity, has various values.
Although the realistic value would be about 0.2 as we have used
for $\soc$, $\rug$, and tubules in the previous papers [7,10],
we scan the parameter space up to $\lambda = 1.4$ to see the
relative stability among possible solutions.

In Fig. 1, we show a set of vectors which specify the tubules on the
honeycomb lattice.  The lattice points in the honeycomb
lattice are labeled by the vector
$(m,n) \equiv m {\bf a} + n {\bf b}$, where ${\bf a}$ and
${\bf b}$ are the unit vectors.  Any structures of tubules can be
produced by connecting the two parallel lines which pass (0,0)
and $(m,n)$ each other and are perpendicular to the vector $(m,n)$.
Hereafter, we use this vector $(m,n)$ to specify the tubules.
When the electron-phonon coupling does not exist, i.e., $\lambda = 0$,
the electronic state of the tubule is classified into metal or
semiconductor depending on the vector.
When the origin of the honeycomb lattice
pattern is superposed with one of the open circles
to make a tubule, the metallic properties will be expected
because of the presence of the Fermi surface.  This case corresponds
to the vectors where $m-n$ is a multiple of three.  If the origin
is superposed with the filled circles, there remains a large gap
of the order of 1eV.  The system will be a semiconductor.  The
similar properties regarding the metallic and semiconducting behaviors
have been discussed in several recent papers [4-6].

When there is a non-zero electron-phonon coupling, several kinds of bond
ordered configurations can be expected.  One of them in the two dimensional
graphite plane is the Kekul\'{e} structure.
The pattern is superposed with the honeycomb
lattice in Fig. 1.  The short and long bonds are indicated
by the thick and normal lines, respectively.  This pattern is
commensurate with the lattice structure for the tubules when $m-n$
is multiples of three in the vector $(m,n)$, because the Kekul\'{e}
structure is the three sub-lattice system.
Therefore, we expect that the Kekul\'{e} structure
is one of the candidates for the most stable solutions.  For other tubules,
the Kekul\'{e} pattern misfits to the boundary condition of the tubule
due to the structural origin.  In contrast, one-dimensional
chain like patterns, where {\sl trans}-polyacetylene--chains
are connected by the long bonds in the transverse direction, can
be realized for any set of $(m,n)$.

We shall explain our strategy of the investigation.  When the
origin $(0,0)$ is combined with $(5,5)$, we obtain an achiral tubule
which has a reflection plane.
In Ref. [5], this tubule has been named as the ``armchair'' fiber.
Both the Kekul\'{e} and chain-like patterns are the candidates for
the stationary solutions.  When we make the tubule $(6,4)$
furthermore, we cut all the rings of ten carbons in tubule
$(5,5)$ and connect the neighboring rings each
other.  The tubule loses the reflection symmetry and becomes chiral.
In this tubule, the Kekul\'{e}
pattern misfits the structure and the chain-like
distortions will be realized.  If we connect next nearest
neighboring rings, we obtain the tubule $(7,3)$.  The tubule becomes
more chiral.  There will be chain-like patterns.
For the tubule (8,2), both the Kekul\'{e} and chain-like structures
become possible again.  Repeating the above procedure further, we
obtain the achiral ``zigzag" fiber at (10,0) [5].  We pursuit changes in
dimerization patterns and electronic energy levels, starting from
the tubule $(5,5)$.

\section{KEKUL\'{E} STRUCTURES IN METALLIC TUBULES}

When Kekul\'{e} structures are commensurate with the honeycomb lattice
pattern, we actually obtain such patterns as one of stationary
solutions.  These are for the cases of the tubules (5,5) and (8,2).
We mainly report the results of the tubule (5,5) with brief comments
for the tubule (8,2).  The system size $N$ is varied within $300
\leq N \leq 600$.  The periodic boundary condition is applied
for the direction of the tubular axis.

For the tubule (5,5), we consider several dimerization patterns, which
are indicated in Fig. 2.  In Fig. 2(a) and (b), the lattice patterns
are Kekul\'{e} type.  In Fig. 2(b), the positions of the short and
long bonds are reversed from those in Fig. 2(a).  Fig. 2(b) contains
hexagons where all the sides are short bonds.  Such
the short bonds are not the double bonds of the ordinary meaning, so
they are denoted by the dashed lines.  Fig. 2(c) and (d) show
the chain-like patterns.  There are several directions of the chains.
For the tubule (5,5), we only consider the directions of chain-like
patterns shown in the figure, because numerical data for the set of the
bond lengths and energy levels for patterns with the other directions
are the same.  There are other possible
patterns where the signs of the bond variables become opposite
from those in Fig. 2(c) and (d).  We have tried to obtain such
kinds of solutions by changing initial bond variables appropriately.
But, we have never obtained them, possibly because
the energy might be much larger or unstable.  Thus, we
shall report the results for the patterns depicted in Fig. 2.
We note that the Kekul\'{e} pattern of Fig. 2(a) has been discussed
in Refs. 3 and 7 but chain-like patterns have not been considered
previously.  We have analyzed possible patterns more extensively.

First, the total energies of the above patterns are shown against the
coupling $\lambda$ in Fig. 3.  The number of carbons is $N=600$. The
energy decreases almost linearly for the strong coupling $\lambda
\sim 1$.  The energies of the patterns (a), (c), and (d) seem
to be almost the same.  However, there are differences larger
than 1eV.  This is clearly reflected to values of the energy gap,
which will be discussed in association with Fig. 4.  The energy of the
pattern (b) is much larger than that of the others. Thus, the
pattern with the reversed alternation of the short and long bonds
is energetically unfavorable.

The energy gap is plotted against $\lambda$ for each pattern in Fig. 4.
The pattern (a) has the widest gap for $0 < \lambda < 0.8$.  This
indicates that the Kekul\'{e} structure is most stable for the realistic
parameter $\lambda \sim 0.2$.  The other patterns are the metastable
states.  When $1.0 < \lambda < 1.4$, the pattern (d) becomes most
stable.  The energy gap $E_g = 6t_0 = 15.0$eV is realized for all the
patterns from (a) to (d).  This is due to the fact that the order
parameters are extraordinarily large at the strong coupling.

Hereafter, we shall discuss properties of the system with $\lambda = 0.2$.
The data for $300 \leq N \leq 600$ are used for the analysis.  This region
of $N$ might yield sufficient data for estimation of electronic and
lattice structures.  We shall concentrate upon properties of the stable
solutions, {\sl i.e.}, the Kekul\'{e} structure of Fig. 2(a).

In Fig. 5, the total energy per site is plotted against $1/N$.
Data points can be fitted well by the parabola curve.  The extrapolated
value to infinite $N$ is -3.9341eV.  This value does not change if we
fit data by a third- or fourth-order polynomial.

Figure 6 shows the energy gap $\eg$ of the tubule (5,5).
The gap varies linearly as a function of $1/N$ due to the one
dimensional nature.  When $N \sim 100$, $\eg$ is of the order of
1eV.  When $N \sim 500$, it becomes of the order of 0.1eV.
The extrapolated value at $N \rightarrow \infty$ is $4.39
\times 10^{-3}$eV.  This is apparently lower than the room temperature.
In addition, we should pay attention to the thermal fluctuation of phonons.
Thus, we can expect nearly metallic behaviors even
at low temperatures.

Figure 7(a) shows the average of the absolute values of the bond
variables, $\langle | y_{i,j} | \rangle$.   This measures the
strength of dimerizations.  The value of $\soc$ is $2.22 \times 10^{-2}$
\AA.  The average $\langle | y_{i,j} | \rangle$ decreases linearly
as a function of $1/N$ due to the one dimensionality.  The extrapolated
value at $N \rightarrow \infty$ is $7.52 \times 10^{-4}$\AA.
This is more than one order of the magnitude smaller than the
observed value in $\soc$ [10] and polyacetylene [8,9].
Figure 7(c) displays the $1/N$ dependence of the bond variables with
the labels of the bonds in Fig. 7(b).  The
length difference between the longest and shortest bonds of $\soc$
is 0.05\AA\ in the present parameters [10].  When $N \sim 500$, the value
becomes about one-tenth of it.  This would be smaller than the maximum value
(about 0.01\AA) of the dimerization strength observable in experiments.
In the narrowest tubules actually present, the maximum
length is about 1000 times of that of the tubule diameter [1].
Such the tubules can be regarded as infinitely long.  The extrapolated
value of the length difference is $2.64 \times 10^{-3}$\AA.
It has been discussed that the widths of the fluctuations of the
bond lengths are of the similar magnitudes (of the order of
10$^{-2}$\AA) in conjugated polymers, graphite plane, and
C$_{60}$ [3,11].  The large fluctuations would make the observation
of the pattern difficult.  Even though the pattern
could not be observed directly,
the Kekul\'{e}-type fluctuations might survive thermal
fluctuations if we look at, for example, the  correlation functions
among lengths of different bonds.

We have also calculated about the tubule (8,2).  We have obtained
the Kekul\'{e} structure as the most stable solution again.
The energy gap $\eg$ and the magnitudes of the bond variables
are not so much different from those of the tubule (5,5), quantitatively.
The number of carbons
arranged around the axis is ten for both tubules.  This indicates that
the electronic and lattice properties are mainly determined by the
diameter of tubules.  They do not sensitively depend on whether
the tubule is chiral or not.

In the previous paper [7], we have looked at the variation of
electronic and lattice structures from $\soc$ and $\rug$ to an
infinitely long tubule (5,5).  Ten more carbons have been inserted
successively for the systematic investigation of changes
in C$_{60}$, C$_{70}$, C$_{80}$, and so on.
The linear $1/N$-dependence is similarly found in the data of $\eg$ and
the bond variables.  We have concluded the Kekul\'{e} pattern with
small dimerization strengths in the lattice structure, too.

\section{CHAIN-LIKE DISTORTIONS IN SEMICONDUCTING TUBULES}

In this section, we discuss about the
tubules, (6,4) and (7,3).  In these tubules, the
Kekul\'{e} pattern is automatically excluded owing to the boundary
condition.  We have obtained only the solutions where
the {\sl trans}-polyacetylene chains are arranged along almost the tubular
axis.  Solutions where chains are oriented in other directions,
are not obtained.  The pattern in the tubule (6,4) is shown in Fig. 8.
Numerical data are reported for $\lambda = 0.2$.

We plot the energy per site vs. $1/N$ in Fig. 9.
It seems that the energy stops decreasing at $N \sim 500$.
Figure 10 shows $\eg$ of the tubule (6,4) as a function of $1/N$.
The energy gap almost saturates at $N \sim 600$.  The large gap
($\sim$ 1eV) remains when $N \rightarrow \infty$.
These saturating properties are due to the fact that there
is a wide gap even in the system with $\lambda = 0$.  The system
is a semiconductor whether there is a dimerization
pattern or not.

Fig. 11(a) shows the variation of the averaged bond variable.
This also saturates at $N \sim 500$.  The value at $N \rightarrow \infty$
is about $1.1 \times 10^{-3}$\AA.  The dimerization strength is
close to that in the nearly-metallic tubules.  This does not depend on
whether there is a gap or not when $\lambda = 0$.   The strength would
be determined mainly by the number of carbons which lie perpendicular
to the tubule axis.  It is ten for all the tubules calculated in this paper.
In Fig. 11(b), we show the bond variables against $1/N$.   The label of
each bond is shown in Fig. 8.  The strength of the dimerization is
very small: the length difference between the shortest
and the longest bonds is about 0.003\AA.  This value is of the
magnitude similar to that found in Fig. 7(c).  The shortest and longest
bonds alternate parallel to the tubular axis.  The bond order is the
strongest in this direction along the ``{\sl trans}-polyacetylene" chains.
This would reflect the one dimensionality.

For the tubule (7,3), we have obtained the same kind of solutions
as the stable solution.
The bond alternation pattern is chain-like.
The energy gap and the strength of the dimerization do not
change so much from those in Figs. 10 and 11.

\section{CONCLUDING REMARKS}

We have investigated the tubules where ten carbons are arranged in the
direction perpendicular to the tubular axis.  We have obtained the
similar strength of dimerizations for the Kekul\'{e} structure and
the chain-like distortion.  The Kekul\'{e} pattern is the most
stable for the metallic tubules, while the chain-like pattern
is realized for the semiconducting tubules.
The strength of the dimerizations is about one order smaller
than the experimentally accessible magnitude.  Therefore, it would
be difficult to observe directly the bond alternation patterns in
the very long tubules.  However, the fluctuations of the phonons
from the classical values might show some correlations which reflect
the Kekul\'{e} or chain-like patterns.

We have estimated the properties of infinitely-long tubules by the
finite-size scaling method.  Certainly, there would be numerical
errors in the extrapolated values.  Calculations using the
wave number space would result in more accurate magnitudes.
However, the present calculations should be valid enough to
estimate the overall magnitudes of the energy gap and the
dimerization strength.

{}~

\noindent
{\bf ACKNOWLEDGEMENTS}\\
The author (K.H.) acknowledges the useful discussion with Dr. K. Yamaji,
Dr. S. Abe, Dr. Y. Asai, and Dr. T. Miyazaki.  He also acknowledges
the hospitality of Department of Physics, University of Sheffield,
United Kingdom, and the financial support for the stay.
A part of this work was done while one
of the author (M.F.) was staying in Department of Physics, Massachusetts
Institute of Technology.  He is grateful for the fruitful collaboration with
Prof. Mildred S. Dresselhaus, Prof. Gene Dresselhaus, and Dr. R. Saito.
Numerical calculations have been performed
on FACOM M-780/20 and M-1800/30 of the Research Information
Processing System, Agency of Industrial Science and Technology, Japan.

\pagebreak
\begin{flushleft}
{\bf REFERENCES}
\end{flushleft}

\noindent
* Electronic mail address: harigaya@etl.go.jp, e9118@jpnaist.bitnet.\\
$[1]$ S. Iijima, Nature {\bf 354}, 56 (1991); T. W. Ebbesen and P. M.
Ajayan, Nature {\bf 358}, 220 (1992).\\
$[2]$ M. Endo, H. Fujiwara, and E. Fukunaga, {\sl Abstract of the
Second C$_{\sl 60}$ Symposium} (Japan Chemical Society, Tokyo, 1992),
pp. 101-104.\\
$[3]$ J. W. Mintmire, B. I. Dunlap, and C. T. White,
Phys. Rev. Lett. {\bf 68}, 631 (1992).\\
$[4]$ N. Hamada, S. Sawada, and A. Oshiyama, Phys. Rev. Lett. {\bf 68},
1579 (1992).\\
$[5]$ R. Saito, M. Fujita, G. Dresselhaus, and M. Dresselhaus,
Phys. Rev. B {\bf 46}, 1804 (1992).\\
$[6]$ K. Tanaka, M. Okada, K. Okahara, and T. Yamabe, Chem. Phys. Lett.
{\bf 193}, 101 (1992).\\
$[7]$ K. Harigaya, Phys. Rev. B {\bf 45}, 12071 (1992).\\
$[8]$ A. J. Heeger, S. Kivelson, J. R. Schrieffer, and W. P. Su,
Rev. Mod. Phys. {\bf 60}, 781 (1988).\\
$[9]$ W. P. Su, J. R. Schrieffer, and A. J. Heeger, Phys. Rev. B
{\bf 22} 2099 (1980).\\
$[10]$ K. Harigaya, Phys. Rev. B {\bf 45}, 13676 (1992).\\
$[11]$ McKenzie and Wilkins, Phys. Rev. Lett. {\bf 69}, 1085 (1992).\\

\pagebreak
\begin{flushleft}
{\bf FIGURE CAPTIONS}
\end{flushleft}

\noindent
Fig. 1. Possible way of making chiral and achiral tubules.
The open and closed circles indicate the metallic and semiconductoring
behaviors of the undimerized system, respectively.  The Kekul\'{e}
structure is superposed on the honeycomb lattice pattern.
The heavy lines indicate the short bonds.

{}~

\noindent
Fig. 2.  Dimerization patterns of the stationary solutions
for the tubule (5,5).  Front and back views are shown by
the normal and thin lines, respectively.  See the text for notations.

{}~

\noindent
Fig. 3.  Total energy per site vs. $\lambda$ for the tubule (5,5).
The system size is $N=600$.  The closed squares are for the
patterns, (a), (c), and (d) (the plots are not desernible),
while the open squares are for (b) in Fig. 2.

{}~

\noindent
Fig. 4.  Energy gap vs. $\lambda$ for the tubule (5,5) with $N=600$.
The closed and open squares are for the patterns (a) and (b), respectively.
The closed and open circles are for (c) and (d).

{}~

\noindent
Fig. 5.  The $1/N$ dependence of the energy per site of the tubule (5,5).

{}~

\noindent
Fig. 6.  The $1/N$ dependence of the energy gap $\eg$ of the tubule (5,5).

{}~

\noindent
Fig. 7.  The $1/N$ dependence of the bond variables of the tubule (5,5).
The averaged bond variable $\langle | y_{i,j} | \rangle$ is shown in (a).
The labels of bonds are shown in (b).  The figure (c) shows the variations
of each bond length.

{}~

\noindent
Fig. 8.  The dimerization pattern of the stationary solution
for the tubule (6,4).

{}~

\noindent
Fig. 9.  The $1/N$ dependence of the energy per site of the tubule (6,4).

{}~

\noindent
Fig. 10.  The $1/N$ dependence of the energy gap $\eg$ of the tubule (6,4).

{}~

\noindent
Fig. 11.  The $1/N$ dependence of the bond variables of the tubule (6,4).
The averaged bond variable $\langle | y_{i,j} | \rangle$ is shown in (a).
The labels of bonds are shown in Fig. 8.  The figure (b) shows the variations
of each bond length.

\end{document}